\documentclass[journal=nalefd,manuscript=letter,layout=twocolumn]{achemso}
\usepackage{chemformula}
\usepackage[T1]{fontenc}
\usepackage{siunitx}
\usepackage{xcolor} 
\usepackage{cuted}
\usepackage{hyperref}
\hypersetup{colorlinks=true,allcolors=black}
\author{Saskia Fiedler}
\affiliation[SDU]
{Center for Nano Optics, University of Southern Denmark, Campusvej 55, DK-5230 Odense M, Denmark}
\alsoaffiliation[equal]
{These two authors contributed equally to the paper}
\author{P.~Elli Stamatopoulou}
\affiliation[SDU]
{Center for Nano Optics, University of Southern Denmark, Campusvej 55, DK-5230 Odense M, Denmark}
\alsoaffiliation[equal]
{These two authors contributed equally to the paper}
\author{Artyom Assadillayev}
\affiliation[DTU]
{Department of Physics, Technical University of Denmark, Fysikvej, DK-2800 Kongens Lyngby, Denmark}
\alsoaffiliation[SDU]
{Center for Nano Optics, University of Southern Denmark, Campusvej 55, DK-5230 Odense M, Denmark}
\author{Christian Wolff}
\affiliation[SDU]
{Center for Nano Optics, University of Southern Denmark, Campusvej 55, DK-5230 Odense M, Denmark}
\author{Hiroshi Sugimoto}
\affiliation[Kobe]
{Department of Electrical and Electronic Engineering, Kobe University, Rokkodai, Nada, Kobe 657-8501, Japan}
\author{Minoru Fujii}
\affiliation[Kobe]
{Department of Electrical and Electronic Engineering, Kobe University, Rokkodai, Nada, Kobe 657-8501, Japan}
\author{N.~Asger Mortensen}
\affiliation[SDU]
{Center for Nano Optics, University of Southern Denmark, Campusvej 55, DK-5230 Odense M, Denmark}
\alsoaffiliation[DIAS]
{Danish Institute for Advanced Study, University of Southern Denmark, Campusvej 55, DK-5230 Odense M, Denmark}
\author{S{\o}ren Raza}
\affiliation[DTU]
{Department of Physics, Technical University of Denmark, Fysikvej, DK-2800 Kongens Lyngby, Denmark}
\email{sraz@dtu.dk}
\author{Christos Tserkezis}
\affiliation[SDU]
{Center for Nano Optics, University of Southern Denmark, Campusvej 55, DK-5230 Odense M, Denmark}
\email{ct@mci.sdu.dk}
\title[Cathodoluminescence particle vs excitation]{Disentangling cathodoluminescence
spectra in nanophotonics: particle eigenmodes \emph{vs} transition radiation }
\begin{document}
\begin{tocentry}

\includegraphics[width=\textwidth]{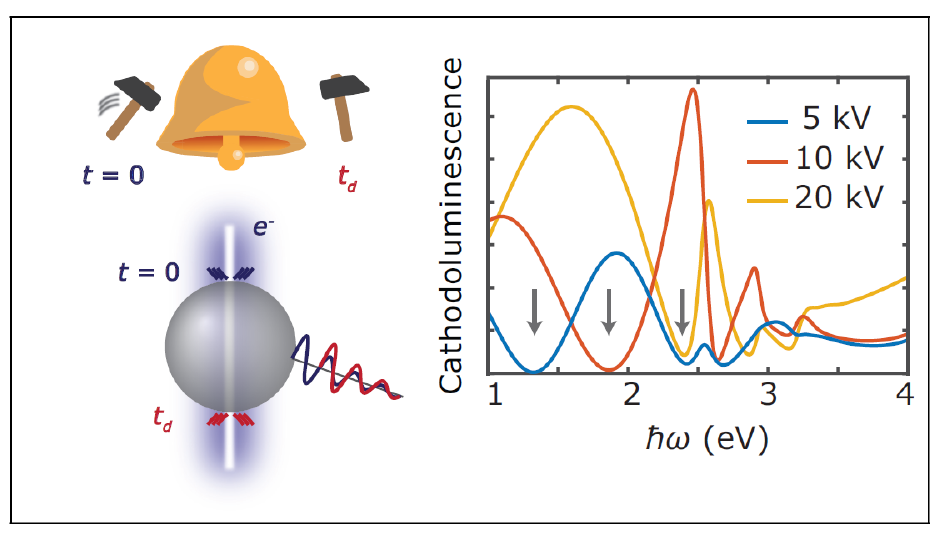}

\end{tocentry}
\begin{abstract}
Cathodoluminescence spectroscopy performed in an electron microscope has
proven a versatile tool for analysing the near- and far-field optical
response of plasmonic and dielectric nanostructures. Nevertheless,
the transition radiation produced by electron impact is often
disregarded in the interpretation of the spectra recorded from
resonant nanoparticles. Here we show, experimentally and theoretically,
that transition radiation can by itself generate distinct resonances
which, depending on the time of flight of the electron beam inside the
particle, can result from constructive or destructive interference in time.
Superimposed on the eigenmodes of the investigated structures, these
resonances can distort the recorded spectrum and lead to potentially erroneous
assignment of modal characters to the spectral features. We develop an
intuitive analogy that helps distinguish between the two contributions.
As an example, we focus on the case of silicon nanospheres, and show
that our analysis facilitates the unambiguous interpretation of
experimental measurements on Mie-resonant nanoparticles.
\end{abstract}
\vspace{1cm}

Electron-based microscopy techniques that harness signals generated
from the excitation of a material by a fast electron beam have proven
essential for exploring the optical properties of
matter~\cite{abajo_rmp82,polman_natmat18}, offering one of the most
efficient platforms for achieving subwavelength-resolution
imaging~\cite{lazar_um106}. They combine the possibility of spatial
resolution optimisation~\cite{abajo_acsphot4} with efficient specimen
excitation~\cite{koh_nn3}, and have met with growing popularity in
quantum- and nano-optics, e.g. 
to study semiconductor nanowires~\cite{yamamoto_apl88}, quantum
dots~\cite{grundmann_prl74,rodt_prb71,mahfoud_jpcl4} or quantum
confinement~\cite{grundmann_jap66,zagonel_nl11}
and to image 
plasmons~\cite{yamamoto_prb64,vesseur_nl7}.
Depending on signal nature and
detection process, electron microscopy and spectroscopy come in
different flavours, such as electron energy-loss spectroscopy
(EELS) and scanning electron microscopy (SEM)~\cite{polman_natmat18}.
Among these, cathodoluminescence (CL) spectroscopy, which collects
the light emitted by the interaction of the electron beam with the
sample, allows direct imaging of optical modes in
plasmonic~\cite{chaturvedi_nn3,sannomiya_nl20} or
dielectric~\cite{matsukata_acsphot6} nanoparticles (NPs), while enabling
detection of optically dark excitations~\cite{koh_nn3}, visualisation of
the local density of optical
states~\cite{abajo_prl100,sapienza_natmat11,zu_nl19},
or even
tomographic reconstruction of the optical near fields~\cite{atre_natnano10}.
These successful endeavours have led to increasingly extended use of CL
spectroscopy as a principal method for analysing photonic nanostructures,
including the characterisation of plasmonic~\cite{matsukata_acsphot5} or
Mie-resonant dielectric NPs~\cite{matsukata_nn15}.

\begin{figure*}[h]
\centering
\includegraphics[width=1\textwidth]{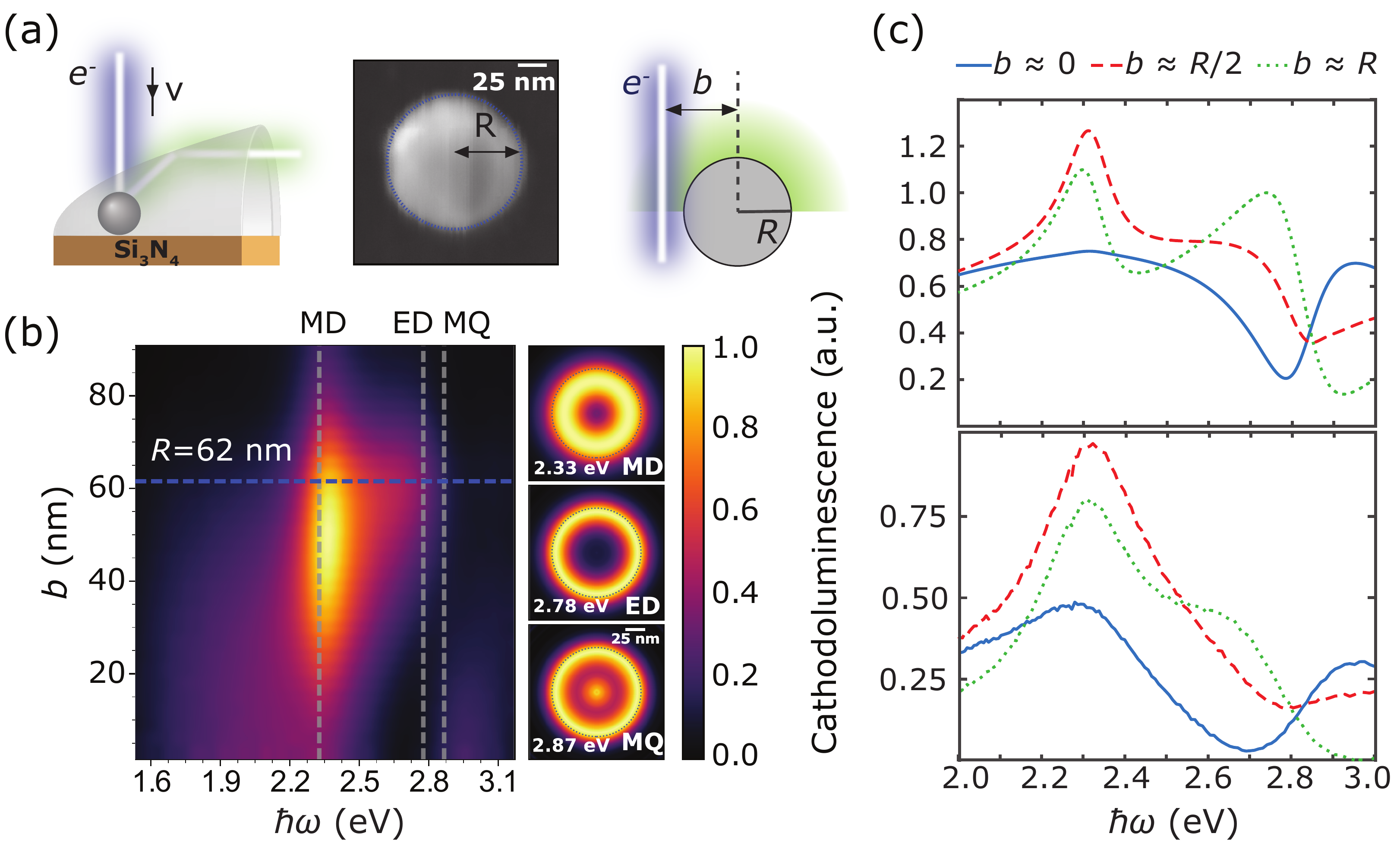}
\caption{
(a) Left: Schematic of the experimental set-up---a Si nanosphere of
radius $R$, placed on a thin Si$_{3}$N$_{4}$ membrane, is excited by
an electron beam. Scattered light is collimated and collected by a
parabolic mirror.
Middle: SEM image of a typical Si NP used in the experiments, with
$R = 62$\;nm.
Right: Sketch of the theoretical description---the electron beam
comes at an impact parameter $b$, and scattered light is collected
at the upper half-space (green-shaded area).
(b) Experimental CL intensity as a function of photon energy
$\hbar\omega$ and impact parameter. Vertical dashed lines denote
the energies of the MD, ED, and MQ resonances predicted by
analytic extinction calculations for the NP of (a). Experimental
CL maps at the MD, ED, and MQ energies are shown on the right-hand
side. In all measurements
here and in panel (c)
the acceleration voltage is $30$\;kV.
(c) Theoretical (upper panel) and experimental (lower panel) CL
spectra for the NP of (a) and three different impact parameters
$b \simeq R$ (green dotted lines), $b \simeq R/2$ (red dashed
lines), and $b \simeq 0$ (blue solid lines).}
\label{fig1}
\end{figure*}

Despite this success, particular care must be taken when interpreting
CL measurements, since the recorded signal can originate from either
excitation of eigenmodes or directly from transition
radiation (TR)~\cite{pogorzelski_pra8,kuttge_prb79,schmidt_nl21}. Here,
we demonstrate how, in the case of electron beams traversing NPs, the
presence of resonances resulting from interfering TR
emanating from the two NP sides with a time delay has the potential to
hinder the unambiguous interpretation of measured spectra in terms of
NP eigenmodes. The two competing mechanisms are, in principle, present
in any CL experiment
---albeit sometimes too weak to observe, depending on the sample---
and can lead to large discrepancies between
anticipated and observed frequencies of NP resonances. We show that
this behaviour is particularly relevant when the electron beam penetrates
the NP near its centre and/or with low velocity, so that its time of flight
in the NP is maximal, and provide an intuitive theoretical description
based on the time delay between the consecutive excitation of radiating
dipoles at the entering and exiting surfaces of the NP. While recent
works have indeed proved CL a powerful tool in the study, design
and monitoring of both dielectric- and plasmonic-based
nanostructures~\cite{abajo_prb59,abuhassan_jap97,vandegroep_optica1,
mcpolin_aom6,peng_prl122,zouros_prb101}, most experiments and analyses
have focused on 
relatively small NPs and
non-penetrating electron beams accelerated at as high
a voltage as possible (so as to excite NP resonances more efficiently
while limiting the electron beam spreading), and the emergence of
additional interference mechanisms for penetrating beams ---and the
possible complications that accompany it--- has remained unexplored.

In what follows, we choose to analyse the emission properties of Si
NPs, because of their high refractive index and low Ohmic
losses~\cite{etxarri_oex19}, the multitude of co-existing modes in the
visible~\cite{staude_nn7,zenin_acsphot7} ---with the field largely
confined inside the NP, thus calling for traversing electron beams---
and the relatively large sizes required for the full glory of all Mie
resonances to unveil itself~\cite{kruk_acsphot4}; the combination of 
these features can significantly pronounce the interference effects
under study. The spectra of Mie-resonant NPs are characterised by
multipoles of both electric and magnetic character~\cite{evlyukhin_nl12},
Fano resonances~\cite{limonov_natphot11}, anapoles~\cite{gurvitz_lpr13},
and bound states in the continuum~\cite{azzam_aom9}, and have enabled
functionalities as diverse as directional light scattering~\cite{fu_natcom4,
person_nl13} and emission~\cite{cihan_natphot12}, directional
couplers~\cite{assadillayev_acsphot8,gulkin_nanophot10}, and Huygens-based
metasurfaces~\cite{jahani_natnano11,staude_natphot11}. All-dielectric
nanodevices are thus proposed as promising alternatives to plasmonics,
with possible applications in biosensing~\cite{etxarri_prb87,yavas_nl17},
nanoantennas~\cite{krasnok_oex18,li_apl107}, slow light~\cite{raza_ol45},
thermo-optic tuning~\cite{assadillayev_nphot10}, ultra-violet interband
plasmonics~\cite{dong_nl19}, fluorescence
control~\cite{schmidt_oex20,stamatopoulou_osac4,sugimoto_acsphot8}, and
Mie-exciton strong-coupling~\cite{tserkezis_prb98,todisco_nanoph9,castellanos_acsphot7,stamatopoulou_prb102}.

CL measurements are performed with the set-up shown schematically in the
left-hand sketch of Figure~\ref{fig1}a (see Methods for details). Si
nanospheres of radius $R$ (prepared in an agglomeration-free colloidal
solution~\cite{sugimoto_aom5}; typical SEM image is shown in the middle
panel of the figure), placed on a thin Si$_{3}$N$_{4}$ membrane, are
exposed to swift electrons travelling at velocity $\mathbf{v}$ (with a
corresponding relativistic factor $\beta = v/c$, where $c$ is the speed
of light in vacuum), for different impact parameters $b$ (see right-hand
schematic of Figure~\ref{fig1}a), corresponding to either non-penetrating
or penetrating electron beams. The colour map of Figure~\ref{fig1}b shows
experimental CL spectra for a relatively small NP ($R = 62$\;nm) ---for
which modal characters should be straightforward to assign--- as a function
of impact parameter. Vertical dashed lines indicate the energies where
different multipoles are predicted by standard Mie theory~\cite{stamatopoulou_osac4}
for this NP size (for extinction spectra and their multipolar decomposition,
see Supporting Information). Two resonances manifest clearly in the spectra
for large impact parameters: a sharp magnetic dipole (MD) at about $\hbar
\omega=2.35$\;eV, and a broader and less intense electric dipole (ED) around
$2.75$\;eV. A magnetic quadrupole (MQ) is identified in the extinction
spectra as a weak contribution to the total extinction at about $2.85$\;eV,
but it is hardly discernible in CL due to losses and its short lifetime.
These features are in good agreement with corresponding dark-field spectra
from literature~\cite{evlyukhin_nl12}, but this agreement becomes worse
for small impact parameters, close to the NP centre. Differences between
theoretical extinction and experimental CL spectra might initially be
attributed to NP shape imperfections and the presence of the thin
substrate~\cite{vandegroep_oex21,fiedler_oex28}, but this turns out to
be only a minor source of deviations, as we discuss below. To better
illustrate the nature of the various resonances, in Figure~\ref{fig1}b
we also show experimental CL maps at the three energies analysed above.

To further analyse the CL spectra, and in order to identify if and where 
theoretical extinction calculations fail to interpret the measurements,
we compare in Figure~\ref{fig1}c experimental spectra (lower panel) with
exact analytic Mie-theory based CL calculations~\cite{matsukata_nn15}
(equivalent to the standard Mie theory for plane-wave excitation, see
Methods),
for three characteristic impact parameters: a grazing electron beam
($b \simeq R$, green dotted lines), a penetrating beam passing halfway
between the edge and centre of the NP ($b \simeq R/2$, red dashed lines),
and one passing exactly through the centre (note that the theoretical
calculations are performed for a small $b \simeq 5$\;nm for convergence
reasons; in any case, this is within the experimental electron-beam diameter).
Different impact parameters are expected to excite different modes with
different weights, while TR-related signals also become relevant when
penetrating beams are considered. Indeed, as the electron beam approaches the
centre, the MD is less efficiently excited for symmetry reasons; in theory,
for $b = 0$ the MD contribution should be exactly zero, but in practice the
electron beam does not follow strictly a straight line (see Supporting
Information for Monte Carlo maps of the electron trajectories). At the same
time, the ED seems to shift in energy as $b$ changes, while even stronger
shifts are exhibited by the dip at about $2.7-2.9$\;eV, identified as an
anapole~\cite{matsukata_acsphot6}. These shifts are not justified by the
poles of the corresponding Mie coefficients (where eigenmodes are located),
and their origin is the main focus of the remainder of this paper.

\begin{figure}[!ht]
\centering
\includegraphics[width=\columnwidth]{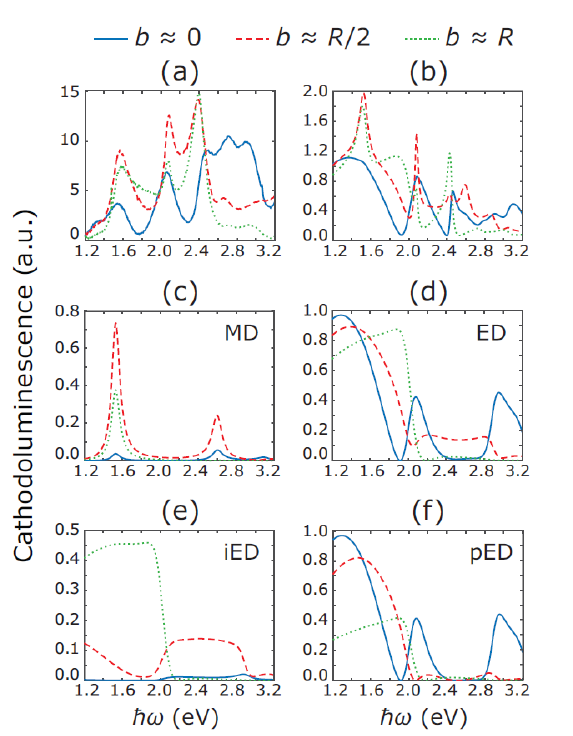}
\caption{
(a) Experimental and (b) theoretical CL spectra for a Si NP with 
$R = 105$\;nm, for three different impact parameter $b \simeq R$ (green
dotted lines), $b \simeq R/2$ (red dashed lines), and $b \simeq 0$ (blue
solid lines).
Theoretical (c)  MD and (d) ED contribution to the CL spectra of (b), for
the same impact parameters. The ED is further analysed into iED and pED
in (e) and (f), respectively.
In all panels the acceleration voltage is $30$\;kV.
}\label{fig2}
\end{figure}

The spectra of Figure~\ref{fig1}c already suggest that a
multipolar decomposition in terms of Mie coefficients might
not be straightforward. To better analyse the situation we
shift in Figure~\ref{fig2} our attention to a larger NP with
$R = 105$\;nm, where higher-order multipoles are expected to
contribute significantly (see corresponding extinction in the
Supporting Information). The MD and ED modes have now moved to
lower energies due to retardation, and additional higher-order
multipoles are clearly visible in the CL spectra at higher
energies (above $\sim 2.5$\;eV), in both experimental
(Figure~\ref{fig2}a) and theoretical (Figure~\ref{fig2}b)
spectra, while the agreement between measurements and
calculations becomes more questionable, especially regarding
the relative peak intensities.
For example, the sharp peak at about $2.4$\;eV
for $b \simeq R$ in Figure~\ref{fig2}b can be attributed to an
electric quadrupole, but it is not clear why its intensity in
Figure~\ref{fig2}a is so much larger.
It is noteworthy that both the resonance peaks and the dips
change intensities and positions as the impact parameter varies.
To assign a multipolar character to such rich spectra, the most
natural approach is to use Mie theory (in its electron-beam
incarnation) to decompose them into independent contributions.
Indeed, in Figure~\ref{fig2}c the MD resonance at about
$1.5$\;eV becomes weaker as the electron-beam trajectory
approaches the NP centre, while remaining practically fixed
in energy, as expected. At the same time, one can observe at
higher energies, around $2.6$\;eV, the emergence of the first
radial MD~\cite{matsukata_acsphot6}. On the other hand, the
ED contribution shown in Figure~\ref{fig2}d is not
straightforward to interpret: it appears as if the main
(first-order radial) ED keeps redshifting away from the
actual pole of the scattering coefficient (see Methods)
as the electron beam approaches the centre, while additional
resonances that cannot be attributed to radial EDs appear
in the energy window $2-3.2$\;eV.
Some kind of plateau is also visible between $2$ and $2.8$\;eV,
which is probably related to the large width of the ED,
combined with what is known as interference structure due
to size in large NPs~\cite{Bohren_Wiley1983}.
This peculiar response is better illustrated in
Figures~\ref{fig2}e-f, where the ED spectrum is further
decomposed into in-plane (iED, plane normal to the electron
beam and parallel to the substrate in the experiment) and
perpendicular (pED, normal to the substrate) EDs ---since
the electric field of the electron beam has non-negligible
components in all directions, it can excite both kinds of
dipolar modes. The most striking feature is the strong shift
(from $2$ to $1.2$\;eV for the lowest-energy resonance),
accompanied by an increase in intensity, and the oscillatory
response of all resonances in the case of the pED shown in
Figure~\ref{fig2}f. There is practically no mechanism that
could shift an ED to such an extent
just by changing the impact parameter
(the pole of the corresponding scattering matrix is at about
$1.9$\;eV). Nevertheless, these oscillations are reminiscent
of the transition radiation manifestation discussed by
Pogorzelski and Yeh~\cite{pogorzelski_pra8}, which
(for $b = 0$) is characterised by a dependence
\begin{equation}\label{eq:sinx}
\Gamma_{\mathrm{CL}} \propto
\frac{\sin(x)}{x}~,
\quad
x = \frac{\omega R}{v}
\end{equation}
for the low (below the Cherenkov limit, as we discuss next)
electron velocities and the dipolar modes of interest here.
Note that the argument $x$ is originally expressed in terms
of the wavenumber (rather than the angular frequency $\omega$)
and the velocity of the electron in the NP, both containing
its refractive index so that it eventually cancels out.
In what follows, we show that this response emerges from the
interference of the TR emitted from the two points where
the electron beam crosses the NP surface
(the same TR that is responsible for the excitation of Mie modes)
with the time delay required for the electron to traverse the NP
from top to bottom, becoming stronger as the electron beam approaches
the NP centre and the electron time of flight increases.

\begin{figure*}[ht]
\centering
\includegraphics[width=\textwidth]{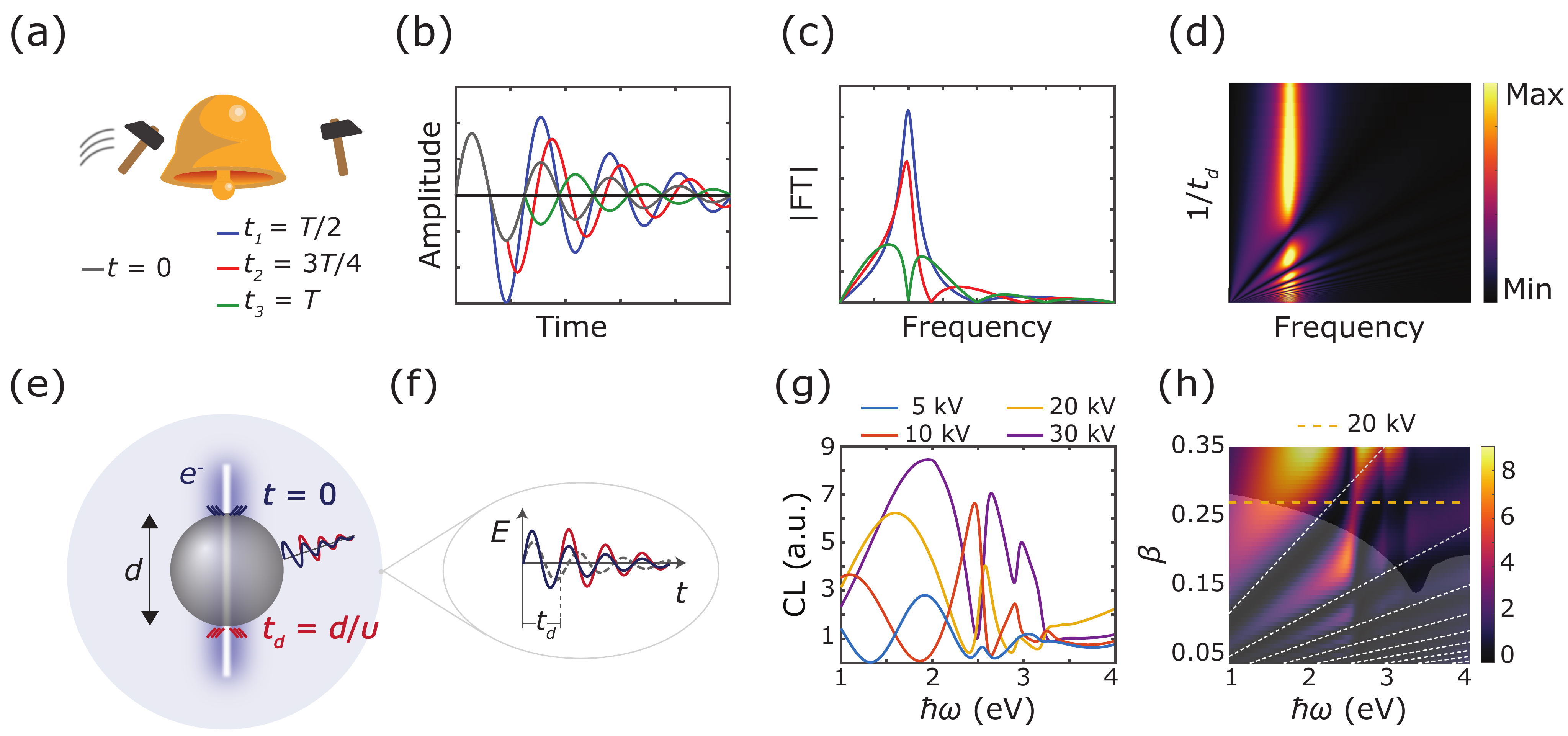}
\caption{(a) A bell hit by two hammers from two opposite sides, with
different time delays $t_{i}$ comparable to the period $T$ of the
generated acoustic waves.
(b) Time evolution of the amplitude of the
acoustic wave generated at $t=0$ (grey line) and the interfering acoustic waves
(blue, red and green lines) produced by the two hammers in the far field. 
(c) FT of the interfering
acoustic waves perceived by an observer. In both (b) and (c) coloured
lines follow the notation of (a).
(d) Colour map of the FT amplitude as a function of frequency and
the inverse of the time delay $t_{\mathrm{d}}$. Destructive
interference conditions follow a linear dependence between
frequency and inverse time delay.
(e) Emergence of the two collapsing dipoles at the upper and
lower surfaces of an NP crossed by an electron beam through
its centre.
(f) Time-domain sketch of the electric far-field amplitude due
to the two excitation events (red and blue lines) and an NP
Mie resonance (grey dashed line).
(g) Calculated CL spectra for a Si sphere with $2R = d = 150$\;nm
for the different acceleration voltages given in the inset, at
$b = 5$\;nm.
(h) Colour map of the calculated CL spectra, as a function of the
electron velocity. Thin white lines serve as guides
to the eye for tracing the interference minima.
The yellow horizontal line indicates the value of $\beta$ that
corresponds to acceleration voltage $20$\;kV, which is analysed
experimentally in Figure~\ref{fig4}. The darker-shaded upper part
of the figure denotes the velocity/energy window where the
Cherenkov condition is met.
}\label{fig3}
\end{figure*}

To obtain an intuitive understanding of the effects discussed above, let us
first focus on the classical analogue analysed in Figures~\ref{fig3}a-d.
Figure~\ref{fig3}a shows a bell which, at time $t = 0$, is hit by a hammer
on its left side, generating an acoustic wave in the far field, with period
$T$. A second hammer hits the bell on the opposite side, with some time delay
$t_{\mathrm{d}}$. The amplitudes of the first
acoustic wave (grey line), and  the three different
superpositions of initial plus delayed wave are
shown as a function of time in Figure~\ref{fig3}b, where we plot three different
time delays for the second signal, corresponding to $T/2$ (blue line), $3T/4$
(red line), and $T$ (green line) ---naturally, all waves decay in time. In
Figure~\ref{fig3}c we plot the Fourier transform (FT) of the total wave that
reaches an observer (in this case listener). The frequency-domain signal
displays an oscillatory behaviour, with zero amplitude emerging whenever
the conditions for destructive interference are met. The amplitude colour
map of Figure~\ref{fig3}d shows, for each time delay, where destructive
interference dips at different frequencies are expected, superimposed over
the natural Lorentzian frequency of the oscillator; a linear dependence
between the inverse of the time delay and the frequency is predicted.

The same concept can equally well apply to the case of NPs excited by
a penetrating electron beam. Thinking of the beam as one
electron approaching the NP at a time, an image charge appears inside
the NP, leading to the formation of an electric dipole. As the electron
approaches the surface, so does the image
charge~\cite{ginzburg_jpUSSR9,goldsmith_pm4,kuttge_prb79} until, at
contact, the two charges collapse, resulting in TR (Figure~\ref{fig3}e).
This collapse also occurs at the bottom side of the particle, with a time
delay $t_{\mathrm{d}} = d/v$ (where $d$ is the electron path length in
the NP). The two dipoles can interfere constructively or destructively,
depending on the time delay of their emergence which, in turn, depends
on the factors that determine the time of flight: $v$ and $d$. 
This response can, in principle, be expected for any NP shape
and can be studied analytically~\cite{abajo_prb69} or
numerically~\cite{losquin_acsphot2}; it should be noted, however,
that the spherical shape stresses the effect, because the electron
beam exits the NP at an interface with air, and collecting the
resulting radiation might be easier.
The corresponding temporal oscillations (red and blue lines), together
with an NP Mie resonance (dashed grey line) are sketched in
Figure~\ref{fig3}f. Exact calculations of CL spectra for different
acceleration voltages (corresponding to different electron velocities)
are shown in Figure~\ref{fig3}g, in the range $5-30$\;kV, which is
experimentally feasible. For low voltages (see, e.g., the blue line for
$5$\;kV) the only recorded signal originates from TR.
As the electron beam is further accelerated, the modes of the NP are more
efficiently excited, leading to CL signals that emerge as a superposition
of NP resonances and interfering TR. This complex interaction is
displayed more clearly in the CL colour map of Figure~\ref{fig3}h where,
in addition to the linear response of the interference dips, one can see
for higher electron velocities slight anticrossings when the NP Mie
resonances are met.
The differently-shaded upper part of Figure~\ref{fig3}h marks the 
velocity/energy window in which the Cherenkov-radiation condition
$v>c/\sqrt{\varepsilon}$ (with $\varepsilon$ being the NP permittivity) is
met. While, by simply looking at this, one could at first think
that Cherenkov radiation plays an important role in the final
form of the measured CL spectra, in the Supporting Information
we show that this is not the case; its only manifestation in
Figure~\ref{fig3}h is the bright feature for $\beta \geq 0.3$
at $\hbar \omega \sim 3$\;eV.
The overall behaviour of the CL spectra suggests thus that CL
minima are extremely sensitive to TR, and care must be taken
so as not to assign to such dips an anapole character before
further verification. Furthermore, the uneven spectral excitation
implies that, even at high acceleration voltages, observed peaks
may be noticeably shifted away from the natural Mie resonances,
which must be kept in mind when interpreting CL measurements.

\begin{figure}[h]
\includegraphics[width=0.8\columnwidth]{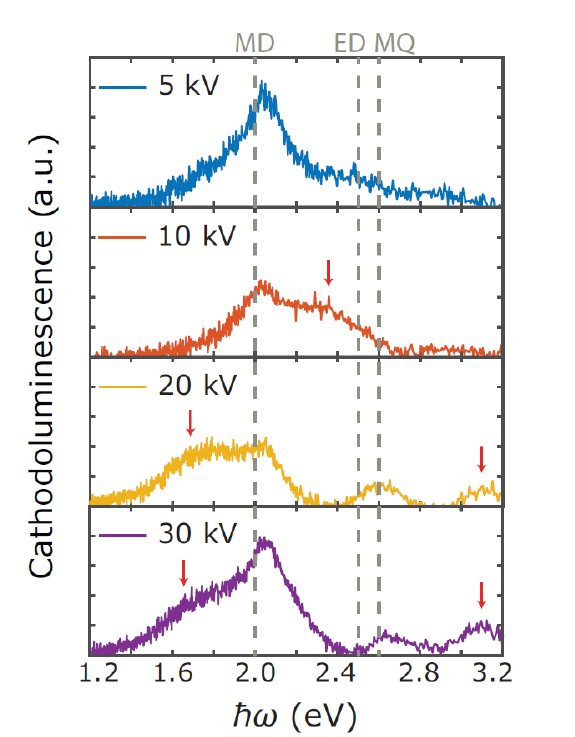}
\caption{
Experimental CL spectra for an electron beam passing through the
centre of a Si nanosphere with $R = 75$\;nm, for acceleration voltages
ranging from $5$\;kV (upper panel) to $30$\;kV (lower panel).
Vertical dashed lines denote the energies
of the MD, ED, and MQ, as predicted by Mie theory. The red markers
indicate spectral features not related to the Mie eigenmodes.
}\label{fig4}
\end{figure}

To verify our explanation, and indeed prove that TR interference
can be observed in experiments with small NPs (as in the case of
thin
films~\cite{yamamoto_jem45,yamamoto_prsla452,stoger_um200},
where Fabry--P\'{e}rot interference emerges due to multiple
reflections) and affect the final spectra, we record in
Figure~\ref{fig4} the CL spectra of a Si sphere with $R = 75$\;nm
at different acceleration voltages, for an electron beam passing
through the NP centre ($b = 0$).
The spectra are affected by the fact that the electron trajectories
are in practice not straight, because the electron mean free path
is just few tens of nm, and the electrons undergo inelastic
scattering before exiting the NP~\cite{lee_jem51} (see Supporting
Information for corresponding Monte Carlo simulations for low and
higher acceleration voltages), thus leading to both inhomogeneous
broadening due to interference events occurring with different
time delays, and to non-negligible contribution from MD terms.
Nevertheless, traces of the oscillations related to TR can
indeed be identified. This is better seen in the spectum for
$20$\;kV, which exhibits four resonances at $1.8$, $2.0$, $2.6$
and $3.1$\;eV, whereas extinction calculations only predict an
MD at about $2$\;eV and an ED/MQ at about $2.5$\;eV; the other
two resonances can be attributed to TR interference or, possibly, 
Cherenkov radiation (in the $3.1$\;eV case), and the energies of
the dips are in excellent agreement with the predictions of
Figure~\ref{fig3}h (see yellow horizontal line in that
figure).
The most remarkable feature is the behaviour of the energy
dip at $2.4 $\;eV for high acceleration voltage ($20-30$\;kV);
as the voltage decreases, this dip transforms into a peak at
the same energy 
(see red markers),
with no correspondence to any NP eigenmode.
Similarly, the shoulder around $1.75$\;eV observed at $30$\;kV,
which disappears for voltages below $10$\;kV, also does not
correspond to any pole of the scattering matrix of the NP. The
above discussion finally sheds more light on the spectra of
Figure~\ref{fig1}c; for external beams, two Mie resonances are
observed at $2.3$\;eV and $2.75$\;eV, while no traces of TR
are present. On the contrary, for penetrating beams, and
particularly when the beam crosses the centre and some of the
Mie modes are inactive for symmetry reasons, it is TR that
dominates the spectra.

\section*{Conclusions}
In summary, we identified TR in CL measurements as a 
possible source of interference with NP resonances, leading to ambiguous,
possibly erroneous, assignment of modal characters to spectral features.
When penetrating electron beams are used, two radiating dipoles are formed
when the electron and its image charge collapse at the two opposite NP surfaces,
with a time delay that is related to the velocity of the electron and the length
of its path inside the NP. These two parameters define whether the interference
of the two dipoles will be constructive, leading to pronounced resonances that
can mask those due to the NP, or destructive, leading to dips reminiscent of
anapoles, but not necessarily of that nature. Care must thus be taken in the
interpretation of CL measurements, especially when characterisation of
NP properties is at question.
To avoid these effects, one can always choose to analyse external
electron beams only, or use as high an acceleration voltage as possible,
while the interference will probably be less pronounced if the shape of
the NP and its positioning are such that the exiting radiating dipole
is formed at the interface with the substrate.

\section*{Methods}

\paragraph*{Cathodoluminescence spectroscopy}
CL spectroscopy is performed in a Tescan Mira3 scanning electron microscope
operated at an acceleration voltage ranging from $5$\;kV to $30$\;kV, current
of $400-470$\;pA, and corresponding spot sizes between $4$\;nm and $15$\;nm.
Light emitted from the sample is collected by a parabolic mirror and analysed
using a Delmic SPARC CL detector equipped with an Andor Newton CCD camera. All
CL spectra are corrected for the system response and the background of the thin
Si$_{3}$N$_{4}$ membrane.
Here, the almost negligible background emission of the substrate,
collected at an unexposed position close to the studied Si NP,
is subtracted from the sample signal and subsequently divided by the
correction curve of the total system response (recorded separately).
All experimental CL spectra in this work correspond to (relative)
photon intensity per wavelength interval, subsequently converted
to energy, taking into account the monochromator dispersion:
$\Delta \lambda = \lambda^{2} \Delta \omega/ (2 \pi c)$, where
$\lambda$ is the wavelength. Therefore, the recorded CL intensity
is multiplied with $\lambda^{2}$. All CL maps are collected with
activated sub-pixel scanning.

In the case of the CL colour map shown in
Figure~\ref{fig1}b, the spectrum is integrated over separate
spatial ranges from $0$ to $1.5 R$ with a step of $0.05 R$. It should be noted that the spectrum which is used for the background subtraction, in this case, is a signal collected by the spectrometer with a blanked electron beam.  Using
a linear interpolation we produce
the data for evenly spaced energy points with the largest energy
separation of $0.0069$\;eV and apply a Gaussian filter with
standard deviation $\sigma = 0.0207$\;eV. Finally, we plot the
resulting spectra as a colour map. The CL maps of different Mie
modes shown in Figure~\ref{fig1}b are plotted for energy-scale
converted CL signal integrated over the wavelength range of $10$\;nm
centred at the resonance energy. We utilise the spherical symmetry
of the Si NPs to radially integrate the signal located at the same
distance from the particle centre and show it as a two-dimensional
map.

\paragraph*{Analytic CL calculation} 
The analytic theory for the interaction of spherical NPs with fast
electron beams has been developed by
Garc{\'i}a de Abajo~\cite{abajo_prb59,abajo_rmp82,matsukata_nn15};
here and in the Supporting Information we only
highlight the key points for completeness.
The angle-integrated photon emission probability
in the interaction of a Si sphere
(described here by the experimental values of Green~\cite{green_semsc92})
with an electron beam passing at impact parameter $b$ with constant
velocity $v$ is described by the expression~\cite{matsukata_nn15}
\begin{equation}\label{eq:G_cl}
\Gamma_\textrm{CL} (\omega) = 
\frac{e^2}{c\pi \varepsilon_0 (\hbar \omega)} \sum_{L} 
\left| \psi_{L}^{\mathrm{ind}} \right|^{2},
\end{equation}
where $e$ is the elementary charge, $c$ is the speed of light in vacuum,
$\varepsilon_{0}$ the vacuum permittivity and $\hbar\omega$
the photon energy. The summation over $L = \left\{P, l, m \right\}$
includes electric ($P = E$) and magnetic ($P = M$) multipoles, characterised
by the angular momentum numbers
$l\geq 1$ and $|m|\leq l$.
Analytic expressions for the
$\psi_{L}^{\mathrm{ind}}$ coefficients are provided in the Supporting
Information;
they describe the expansion of the field of a moving electron
into spherical waves, for the two different regions (inside and outside
the NP).

\begin{acknowledgement}
S.~R. is a Sapere Aude research leader supported by Independent Research
Fund Denmark (Grant No. 7026-00117B).
N.~A.~M. is a VILLUM Investigator supported by VILLUM FONDEN
(Grant No.~16498).
C.~W. acknowledges funding from a MULTIPLY fellowship under the
Marie Sk\l{}odowska-Curie COFUND Action (Grant agreement No.~713694).
P.~E.~S. is the recipient of the Zonta Denmark's Scholarship for
female PhD students in Science and Technology 2021.
\end{acknowledgement}

\begin{suppinfo}

Details about the analytic CL calculation; Mie-theory extinction spectra;
Monte Carlo simulations of electron trajectories, Sample degradation test.

\end{suppinfo}

\bibliography{Si_CL_refs.bib}

\end{document}